# A Three-Terminal Spin-Wave Device for Logic Applications

Yina Wu, Mingqiang Bao, Alexander Khitun, Ji-Young Kim, Augustin Hong, and Kang L Wang

Device Research Laboratory, Electrical Engineering Department,

University of California Los Angeles, Los Angeles, CA, 90095-1594

FCRP Focus Center on Functional Engineered Nano Architectonics (FENA),

Nanoelectronics Research Initiative - The Western Institute of Nanoelectronics (WIN)

**Abstract**

We demonstrate a three-terminal spin wave-based device utilizing spin wave interference. The device consists of three coplanar transmission lines inductively coupled to the 100nm thick CoFe film. Two spin wave signals are excited by microwave fields produced by electric current in two sets of lines, and the output signal is detected by the third set. The initial phases of the spin wave signals are controlled by the direction of the current flow in the excitation set of lines. Experimental data show prominent output signal modulation as a function of the relative phase (in-phase and out-of phase) of two input signals. The micrometer scale device operates in the GHz frequency range and at room temperature. Our experimental results show that spin-wave devices exploiting spin wave interference may be scaled to micrometer and nanometer scales for potential logic circuit application.

The continued scaling of the feature size of Complementary Metal Oxide Semiconductor (CMOS) devices, as it is known today, will likely meet major challenges end due to the power dissipation, manufacturing complexities, and eventually the feature size will approach the tunneling limit. Spintronics came to the stage to exploit electron spin beyond electron charge as an information carrier [1]. Several spin based semiconductor logic devices have been proposed [2-4]. Spin wave bus utilizes spin waves as a physical mechanism for information transmission without the transport of carriers (electrons and holes) [5]. The key advantage is that information transmission is accomplished without a charge transfer. Prototype spin-wave logic device has been experimentally demonstrated [6]. In addition, the functionality of spin-wave logic exclusive-not-OR and not-AND gates has also been realized [7]. In the above cited works[6, 7], a Mach-Zehnder-type spin-wave interferometer was used to achieve the output power modulation as a result of the interference of two spin waves. The relative phases of two spin waves were controlled by the magnetic fields applied separately to the interferometer paths. The magnetic fields were produced by electric currents through the copper wires fabricated beyond the waveguides. These works experimentally demonstrated the possibility of using spin wave interference to achieve logic functionality in millimeter scale at room temperature. In this letter, we report a three-terminal spin wave device, where the phase shift between two input spin wave signals is achieved by controlling the direction of current flow in two sets of excitation lines. This approach makes possible to reduce the length of the device to micrometer and nanometer scale and provides a route to scalable spin-wave based logic circuits [8, 9].

In Fig.1, we show the schematic of the prototype spin wave-based device, and the device layout used for the experiments. The spin wave device was fabricated on a silicon substrate: a layer of $SiO_2$ thin film was first deposited by Chemical Vapor Deposition (CVD) before a 100 nm thick ferromagnetic $Co_{30}Fe_{70}$ thin film was sputtered on the top of the $SiO_2$ film using a high vacuum rf-sputtering system. The film exhibited a saturation magnetization ($B_s$) of ~2.2 T. Then, a further 300 nm layer of CVD $SiO_2$ was deposited on the $Co_{30}Fe_{70}$ film. After that, 500 nm thick gold electrodes were deposited on the top of the CVD $SiO_2$ film and patterned by low temperature plasma etching. The size of the $Co_{30}Fe_{70}$ thin film is 1.0 mm × 0.72 mm. There is no ferromagnetic film under the contact pad region.

To study spin wave transport in a 100 nm thick $Co_{30}Fe_{70}$ film, we use the propagating spin wave spectroscopy (PSWS) technique that was well described in [10-12]. The device under test (DUT in Fig.1(a)) has three set of asymmetric coplanar strip lines (ACPSL). The spin waves are excited by the first and third edge ACPSLs, and detected by the center ACPSL. For a tangentially magnetized slab, there are two modes of spin waves that can be excited depending on the relative orientation of the magnetization M with respect to the in-plane wave vector k: magnetostatic backward volume wave (MSBVW) for M∥k; and magnetostatic surface wave (MSSW) for M⊥k [13]. We restrict our experiment to the MSSW configuration, as the MSSW mode has a higher group velocity and gives relatively strong signal in the particular frequency range. The dispersion relation for the MSSW mode propagation with a small fixed magnetic field ($H+ H_k <<M_s$) is given in the Damon-Eschbach model [9],

$$f(k,H) = \frac{\gamma\mu_0}{2\pi}\sqrt{M_s(H_k + H) + (\frac{1}{2}M_s)^2[1-\exp(-2kd)]} \quad (1)$$

where $\gamma = \frac{g\mu_B}{\hbar}$ is the gyromagnetic ratio, g is the gyromagnetic splitting factor, $\mu_B$ is the Bohr magneton, $\hbar$ is Plank's constant, $\mu_0$ is the permeability of free space, $H_k$ is the in-plane anisotropy field, $M_s$ is the saturation magnetization, and $d$ is the thickness of CoFe thin film. Eq. (1) defines the frequency window (the lowest frequency $f(0,H) = \frac{\gamma\mu_0}{2\pi}\sqrt{M_s(H_k + H)}$, and the highest frequency $f(\infty,H) = \frac{\gamma\mu_0}{2\pi}(H + H_k + \frac{1}{2}M_s)$) in which magnetostatic spin waves can propagate in a ferromagnetic thin film at a given external magnetic field $H$.

In Fig.2, we present experimental data showing the inductive voltage measured at the central ACPSL at different values of the external magnetic field strength (up to 650 Oe) at the excitation frequency of 1.5 GHz. The data plotted in Fig.2 is a result of subtraction. In order to exclude the direct magnetic coupling among the lines via stray field, we normalize the values obtained for all magnetic field strength $H$ to that obtained at 650 Oe, at which there is no spin wave propagation through the slab. The red and blue curves depict the output voltage produced by two input spin wave signals, which are for the in-phase ($\Delta\phi=0$) and out of phase ($\Delta\phi=\pi$) cases, respectively. The phase shift between the spin wave signals is defined by the direction of the current flow in the ACPLs. The phase difference is zero if the direction of the current flow is the same (clockwise or counter-clockwise wise) in both input sets of lines. If the directions in the excitation sets of lines are such as one loop is clockwise and the other is counter-clockwise, the spin wave signals receive a π relative phase difference. As one can see from Fig.2, there is a prominent output voltage difference (up to 2 times) between in-phase and out of phase signals in the magnetic field window from 10 Oe to 150 Oe. Here, we stress that the observed output power modulation is achieved by the control of the relative *phases* of two input signals.

In Fig.3, we present experimental data obtained at 1.0 GHz, 2.0 GHz and 3.0 GHz for in-phase and out of phase cases (red and blue curves, respectively). All measurements are carried out at room temperature. These data illustrate some common trends in the observed spectra: (i) the amplitude of the output signal decreases to zero at high external magnetic field, (ii) the difference between the in-phase and out of phase signals can be observed only in a certain magnetic field window as the magnetostatic spin waves in a given slab configuration can propagate only in a certain frequency/magnetic field range as seen in Eq.(1). The out of phase signal has lower amplitude than that of the in-phase signal. However, the amplitude of the out of phase signal is not zero, as one might expect for the destructive interference. This fact can be attributed to the finite size of the receiver and asymmetric configuration of the ACPS lines, which has a narrow bandwidth.

The main difference and, potentially, an advantage of the proposed spin wave device over the devices presented in [6] and [7] are discussed below. The minimum length of the Mach-Zehnder-type spin-wave interferometer used in the above-cited works is limited by the length required to achieve a significant (about π) phase shift. The latter translates into the need of high modulation field or/and a relatively long path length. In contrast, in our proposed design, the π phase shift between two spin wave signals can be easily achieved by the control of the direction of current flow in the excitation line. Thus, the distance between the transducer and receiver is no longer limited by the phase accumulation length and can be scaled down to the spin wave wavelength. However, there is one common shortcoming inherent to all spin wave based devices demonstrated so far, i.e. lack of signal gain. In order to build a large cascade logic circuit, one has to provide a signal amplification to compensate the signal attenuation due to the spin wave damping. It will require a special device to amplify the amplitude of the spin wave signal.

In conclusion, we demonstrated a room temperature working three-terminal prototype device utilizing spin waves interference for output voltage modulation. The relative phases of two spin wave signals were controlled by the direction of the current flow in the excitation set of lines. The maximum signal modulation of a factor of two was observed at 1.5 GHz frequency with a 50 Oe magnetic field in the MSSW configuration. The obtained data showed the possibility of using spin waves for implementation of scalable logic devices on a silicon platform. The demonstrated prototype three-terminal device is a step toward the spin wave-based logic circuitry. The utilization of wave properties such as superposition and interference may open a new horizon for logic devices alternative to current CMOS-based circuits.

Acknowledgments

We would like to thank Dr. S. Wang and D.W. Lee (Stanford University) for CoFe deposition and Dr. Ajey Jacob from Intel for valuable discussion. The work was supported in part by the Focus Center Research Program (FCRP director: Dr. Betsy Weitzman) Center of Functional Engineered Nano Architectonics (FENA), and by the Nanoelectronics Research Initiative (NRI Director: Dr. Jeff Welser) - The Western Institute of Nanoelectronics (WIN).

Figure Captions:

Fig1 (a) Schematics of the prototype three-terminal device. The structure from the bottom to top consists of a silicon substrate, a 100 nm thick $Co_{30}Fe_{70}$ film, and a 300 nm thick silicon dioxide layer. There are three sets of ACPS lines on top of the structure. Two edge sets are used to excite two spin wave signals and the middle one is used to detect the inductive voltage produced by two spin waves. (b) Photo of the fabricated device. The red circle depicts the three sets of ACPS lines. (c) Magnified view of the ACPS lines. The dimensions of the ACPS lines: Wg = 9 um, Ws = 2 um, Wt = 4 um, and gap between transducer and receiver g = 2 um. Both edge sets of ACPSLs are used as the input transducers and the center one is used as the output receiver. The arrows show the directions of spin wave propagation from the two edge sets to the center line.

Fig.2 (a) Schematics to illustrate phase difference between two input spin wave signals. The initial phase difference between the input signals is defined by the direction of the current flow in the lines. The phase difference is zero if the direction of current flow is the same (either clockwise or counter-clockwise) in both lines. If the directions of the currents in the lines are opposite, the two spin wave signals have a π relative phase difference. b) Experimental data show the output inductive voltage for the in-phase (red curve) and out of phase (blue curve), respectively. The excitation frequency used is 1.5 GHz at room temperature. The external magnetic field strength was varied from –650 Oe to 650 Oe. The sign of the magnetic field depicts the direction (along or opposite to x axis shown in Fig.1).

Fig.3 Experimental data obtained for excitation frequency of 1 GHz, 2 GHz, and 3 GHz. The external magnetic field strength is varied from –650 Oe to 650 Oe. The sign of the magnetic field is shown along or opposite to X axis in Fig.1.

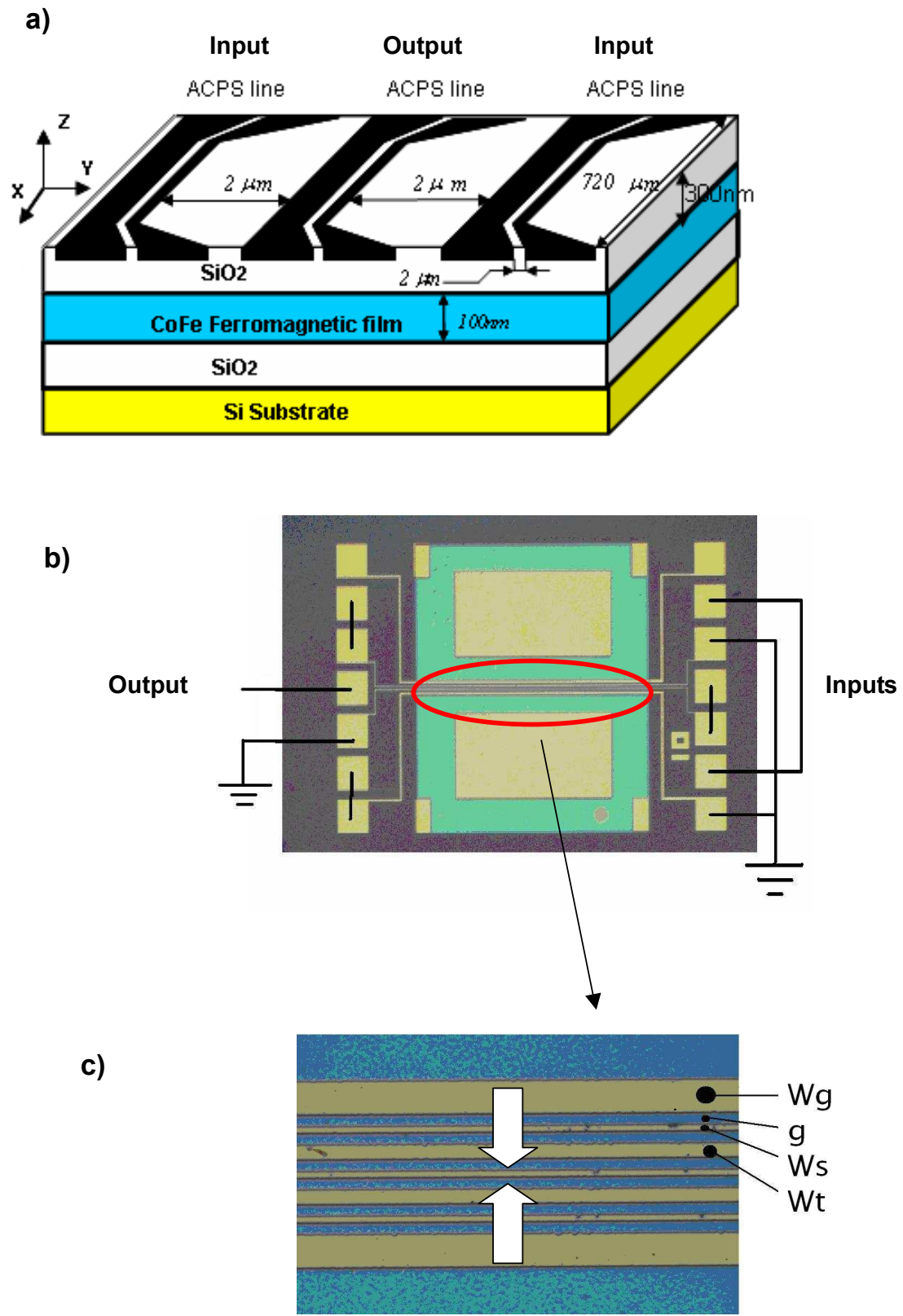

**Fig.1**

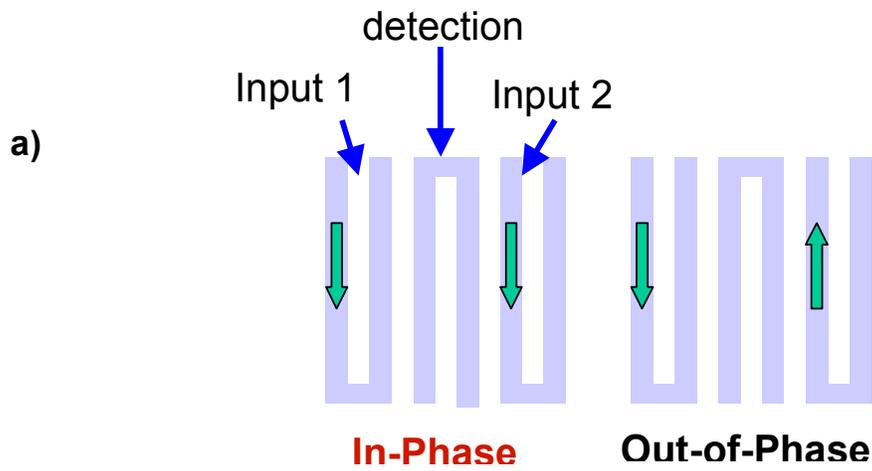

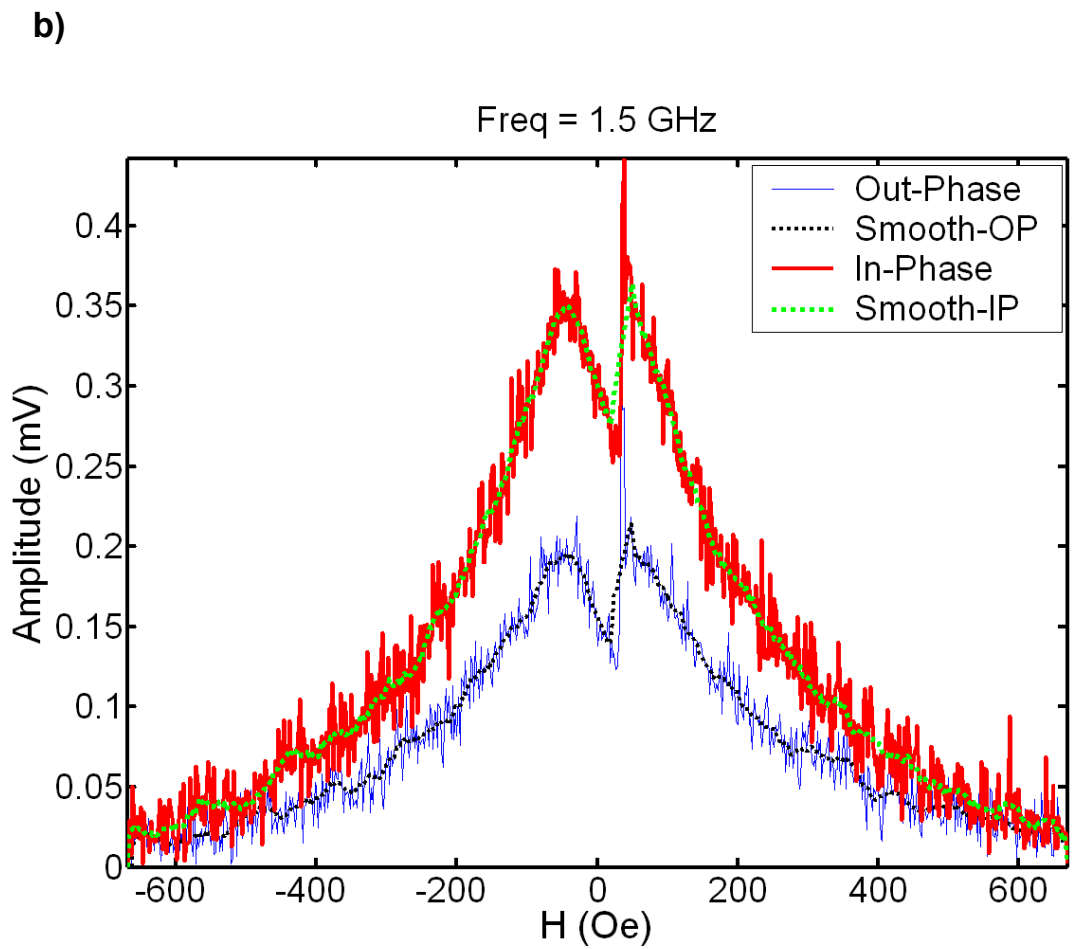

**Fig.2**

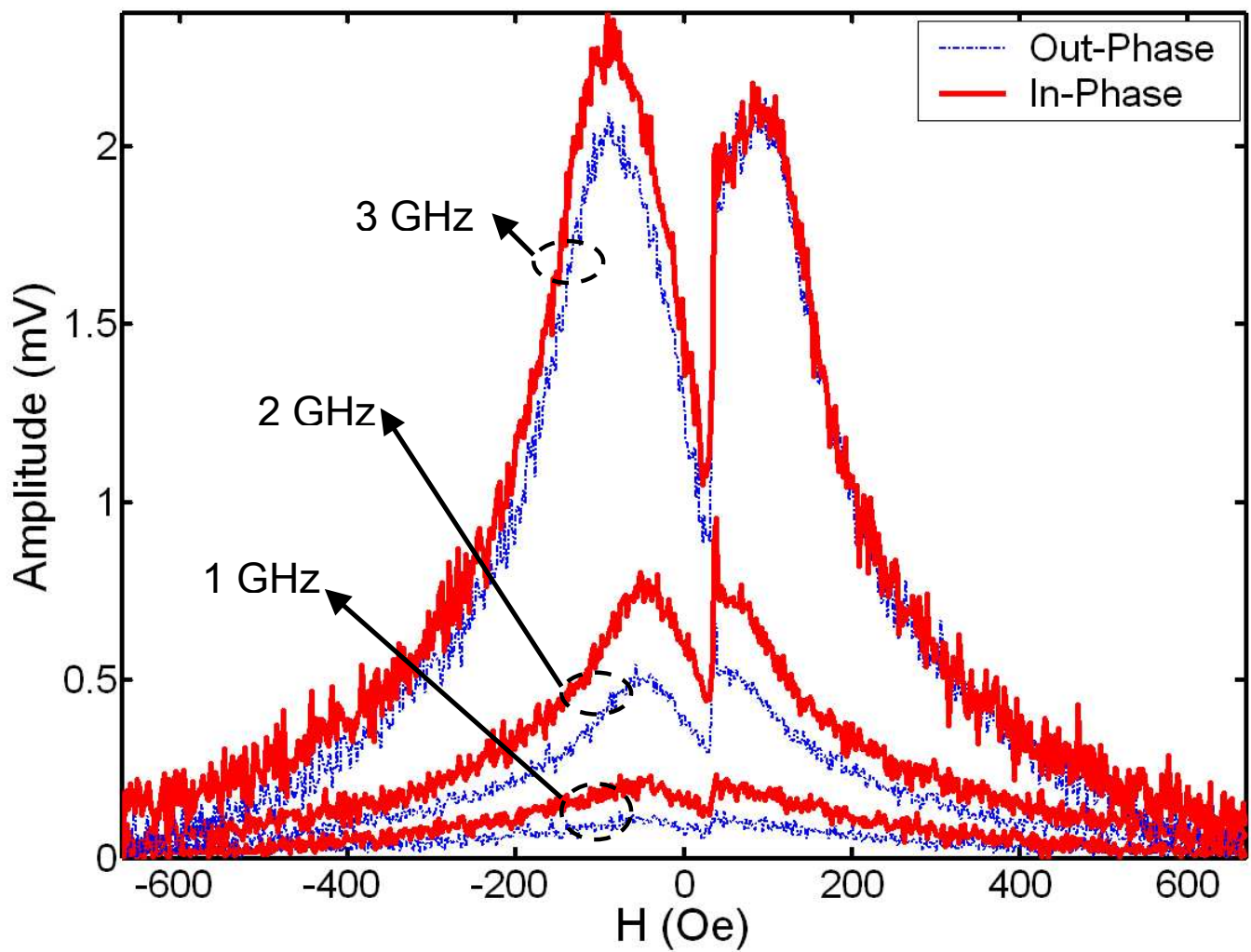

**Fig.3**